\documentclass[amsmath,amssymb,aps,pra,reprint,groupedaddress,showpacs]{revtex4-1}

\usepackage{multirow}
\usepackage{verbatim}
\usepackage{color,graphicx}

\newcommand{\Proof}{\noindent\textbf{Proof.}\quad}
\newcommand{\qed}{\hfill$\Box$}

\newtheorem{theorem}{Theorem}
\newtheorem{lemma}[theorem]{Lemma}

\newtheorem{proposition}[theorem]{Proposition}

\begin{document}

\title{Ability of stabilizer quantum error correction to protect itself from its own imperfection}

\author{Yuichiro Fujiwara}
\email[]{yuichiro.fujiwara@caltech.edu}
\affiliation{Division of Physics, Mathematics and Astronomy, California Institute of Technology, MC 253-37, Pasadena, California 91125, USA}

\date{\today}

\begin{abstract}
The theory of stabilizer quantum error correction allows us to actively stabilize quantum states and simulate ideal quantum operations in a noisy environment.
It is critical is to correctly diagnose noise from its syndrome and nullify it accordingly.
However, hardware that performs quantum error correction itself is inevitably imperfect in practice.
Here, we show that stabilizer codes possess a built-in capability
of correcting errors not only on quantum information but also on faulty syndromes extracted by themselves.
Shor's syndrome extraction for fault-tolerant quantum computation is naturally improved.
This opens a path to realizing the potential of stabilizer quantum error correction
hidden within an innocent looking choice of generators and stabilizer operators that have been deemed redundant.
\end{abstract}


\maketitle

\section{Introduction}
Quantum error correction plays the central role in stabilizing inevitably fragile quantum states and
simulating perfect quantum operations in a noisy environment \cite{Nielsen:2000,Lidar:2013}.
A critical problem the theory of quantum error correction faces
is that quantum gates that perform error correction themselves are faulty in practice.
Therefore, we must build our quantum information processing device on an architecture that does not fall apart even if all components,
including those responsible for quantum error correction, are imperfect.
Such robust architectures are \textit{fault-tolerant}.

Fault tolerance is of particular significance
because the theory of quantum error-correcting codes typically assumes perfect execution of error correction procedures.
For instance, \textit{stabilizer codes} \cite{Gottesman:1996,Calderbank:1998} are the most extensively studied quantum error-correcting codes
that form a very general and important class.
Quantum error correction via stabilizer codes diagnoses noise by extracting \textit{syndromes},
which indirectly tell us how quantum information may have been degraded.
Because the conventional theory of stabilizer codes does not provide protection of syndromes on its own,
it has been considered that external help is required to achieve robust syndrome extraction.

The primary purpose of this work is to show that, contrary to this conventional wisdom,
stabilizer codes have a built-in capability of correcting faulty syndromes on their own.
In other words, the theory of quantum error-correcting codes is shown to be able to reduce the burden on the shoulders of a fault-tolerant architecture.
Aspects of quantum error correction that have been considered irrelevant or redundant play a key role in realizing the full potential of stabilizer codes.

It should be noted, however, that our findings are not a replacement for fault-tolerant syndrome extraction.
Rather, the innate ability of stabilizer codes we will reveal augments the existing framework.

There are primarily three known fault-tolerant methods for quantum syndrome extraction,
which were discovered by Shor \cite{Shor:1996}, Steane \cite{Steane:1997}, and Knill \cite{Knill:2005,Knill:2005a} respectively.
The simplest and most general is Shor's method (see also \cite{DiVincenzo:1996}).
Unlike the other two schemes, it does not require complicated quantum states, which makes implementation easier.
Moreover, it works for all stabilizer codes.

Fortunately, Shor's fault-tolerant method is particularly suited for exploiting the innate ability of stabilizer codes.
Roughly speaking, the central idea of Shor's robust syndrome extraction is to repeat the same set of measurements for syndrome extraction in a safe manner,
so that each repetition increases confidence that the observed syndrome is correct
while avoiding propagation of the effects of errors on quantum information and possible failure of quantum circuits.
Our observations naturally extend Shor's method and help reduce the required number of measurements
by carefully choosing which measurement should be performed.

The next section provides a brief review of stabilizer error correction.
Section \ref{sec:main} explains our main idea for robust syndrome extraction.
Its implication in the context of fault tolerance and a main remaining problem beyond the scope of this work are discussed in Section \ref{sec:ft}.
Section \ref{sc:cr} concludes this paper with further remarks.

\section{Stabilizer codes}\label{sec:review}
We briefly review the theory of stabilizer quantum error correction.
For a more comprehensive introduction, we refer the reader to \cite{Nielsen:2000,Lidar:2013}.

Take the Pauli group $\mathcal{P}$ over $n$ qubits, which consists of the $n$-fold tensor products of
Pauli operators $X$, $Y$, and $Z$ as well as the trivial operator $I$ with overall factors $i^\lambda$, where $\lambda \in \{0,1,2,3\}$.
The \textit{weight} $\operatorname{wt}(E)$ of $E\in\mathcal{P}$ is the number of nontrivial operators in its $n$ factors.
All quantum error-correcting codes we consider are realized
as $2^k$-dimensional subspaces of the full $2^n$-dimensional Hilbert space $(\mathbb{C}^2)^{\otimes n}$,
so that $k$ logical qubits are encoded into $n$ physical qubits, which we call \textit{data qubits}.
In particular, an $[[n,k,d]]$ \textit{stabilizer code} is the unique $2^k$-dimensional subspace $\mathcal{H}_\mathcal{S}$ stabilized by
an abelian subgroup $\mathcal{S}$ of $\mathcal{P}$ with $-I^{\otimes n} \not\in \mathcal{S}$ generated by $n-k$ independent operators
such that $\min\{\operatorname{wt}(C) \mid C\in\mathcal{C}_\mathcal{S}\setminus\mathcal{S}\} = d$,
where $\mathcal{C}_\mathcal{S} = \{E \in \mathcal{P} \mid ES = SE \text{ for all } S \in \mathcal{S}\}$.
The group $\mathcal{S}$ is the \textit{stabilizer} of $\mathcal{H}_\mathcal{S}$.
Each $S \in \mathcal{S}$ is a \textit{stabilizer operator}.
The minimum weight $d_p=\min\{\operatorname{wt}(C) \mid C\in\mathcal{C}_\mathcal{S}\setminus\{I\}\}$ is the \textit{pure distance}.
The stabilizer code is \textit{degenerate} if $d>d_p$ and \textit{nondegenerate} otherwise.

All standard error correction schemes for stabilizer codes involve \textit{discretization}, which collapses an arbitrary error into some operator $E\in\mathcal{P}$ \cite{Knill:1997a}.
Thus, without loss of generality, we assume that noise is tensor products of Pauli operators.
In this setting, an $[[n,k,d]]$ stabilizer code can correct any error $E\in\mathcal{P}$ with $\operatorname{wt}(E)\leq\lfloor(d-1)/2\rfloor$.

The \textit{syndrome bit} $s_i(E)$ of $E$ by the $i$th stabilizer operator $S_i$ is $0$ if $E$ and $S_i$ commute and $1$ otherwise.
The vector $(s_0(E),\dots,s_{2^{n-k}-1}(E))$ is the \textit{full syndrome} of $E$.
Note that each syndrome bit is a linear combination of those given by the generators $G\in\mathcal{G}$, where $\mathcal{S}=\langle\mathcal{G}\rangle$.
Thus, $n-k$ independent syndrome bits contain as much information about $E$ as the full syndrome.

We illustrate how $n-k$ syndrome bits reveal which error occurred by using the \textit{perfect} $5$-\textit{qubit code} \cite{Laflamme:1996,Bennett:1996} as an example.
The following four operators generate the stabilizer of a $2$-dimensional subspace of $(\mathbb{C}^2)^{\otimes 5}$:
\begin{align*}
S_0 &= XZZXI, & S_1 &= IXZZX,\\
S_2 &= XIXZZ, & S_3 &= ZXIXZ,
\end{align*}
where the symbol ${\otimes}$ for the tensor product is omitted.
Any nontrivial Pauli operator on one qubit is identified by its syndrome as shown in Table \ref{tbl:syndrome5qubit}.
\begin{table}[h!t]\caption{Syndromes by the perfect $5$-qubit code.\label{tbl:syndrome5qubit}}
\begin{ruledtabular}
\begin{tabular}{cc|cc}
Error&$(s_0, s_1, s_2, s_3)$&Error&$(s_0, s_1, s_2, s_3)$\\\hline
No error&$(0, 0, 0, 0)$&$IIYII$&$(1, 1, 1, 0)$\\
$XIIII$&$(0, 0, 0, 1)$&$IIIYI$&$(1, 1, 1, 1)$\\
$IXIII$&$(1, 0, 0, 0)$&$IIIIY$&$(0, 1, 1, 1)$\\
$IIXII$&$(1, 1, 0, 0)$&$ZIIII$&$(1, 0, 1, 0)$\\
$IIIXI$&$(0, 1, 1, 0)$&$IZIII$&$(0, 1, 0, 1)$\\
$IIIIX$&$(0, 0, 1, 1)$&$IIZII$&$(0, 0, 1, 0)$\\
$YIIII$&$(1, 0, 1, 1)$&$IIIZI$&$(1, 0, 0, 1)$\\
$IYIII$&$(1, 1, 0, 1)$&$IIIIZ$&$(0, 1, 0, 0)$\\
\end{tabular}
\end{ruledtabular}
\end{table}
Indeed, it can be checked that these stabilizer operators define a $[[5,1,3]]$ code.
It is \textit{perfect} because all $2^{n-k}$ possible patterns of syndromes are used up to distinguish single errors and no error from each other.

\section{Correcting imperfect syndromes by stabilizer codes themselves}\label{sec:main}
The above theory relies on the assumption that all syndrome bits are noiseless.
However, it is plausible that errors occur on syndromes, potentially causing $1$ to be flipped to $0$ or vice versa.
Possible causes include imperfect ancilla qubits holding syndromes and faulty measurements of stabilizer operators.
Shor's syndrome extraction handles this kind of error by repeating the same syndrome measurements until enough confidence is gained.
We generalize this strategy.

To illustrate our key insight as plainly as possible, we focus for the moment on how many data qubits and syndrome bits are allowed to be erroneous.
This view is reasonable if no error occurs on data qubits during syndrome extraction.
This error model was very recently studied in \cite{Ashikhmin:2014} as well in the context of robust syndrome extraction
primarily with implementation via trapped ions in mind.

\subsection{Global single error correction}\label{subsec:single}
Now, using the same single-error-correcting $5$-qubit code as before, let us assume that
one of the five data qubits or the four syndrome bits is erroneous after syndrome extraction.
Since the perfect code already uses up all $2^4=16$ different syndromes,
at first glance the stabilizer does not seem to possess error correction power for syndrome bits on its own.
In fact, if the syndrome bit $s_3$ is flipped when there is no error on the data qubits,
we end up with the erroneous syndrome $(0,0,0,1)$, which is the same as the correct syndrome of $X$ acting on the first qubit.
Fortunately, the reality is not as pessimistic.

Take stabilizer operator $S_4 = \prod_{i=0}^3{S}_i$.
The conventional theory of quantum error correction does not use $S_4$ because it is considered ``redundant.''
However, as shown in Table \ref{tbl:extendedsyndrome5qubit},
joining $S_4$ allows for distinguishing all possible single errors including those on syndrome bits.
\begin{table}[h!t]\caption{Syndromes with a redundant stabilizer operator.\label{tbl:extendedsyndrome5qubit}}
\begin{ruledtabular}
\begin{tabular}{cc|cc}
Error&$(s_0, s_1, s_2, s_3,s_4)$&Error&$(s_0, s_1, s_2, s_3,s_4)$\\\hline
No error&$(0, 0, 0, 0, 0)$&$ZIIII$&$(1, 0, 1, 0, 0)$\\
$XIIII$&$(0, 0, 0, 1, 1)$&$IZIII$&$(0, 1, 0, 1, 0)$\\
$IXIII$&$(1, 0, 0, 0, 1)$&$IIZII$&$(0, 0, 1, 0, 1)$\\
$IIXII$&$(1, 1, 0, 0, 0)$&$IIIZI$&$(1, 0, 0, 1, 0)$\\
$IIIXI$&$(0, 1, 1, 0, 0)$&$IIIIZ$&$(0, 1, 0, 0, 1)$\\
$IIIIX$&$(0, 0, 1, 1, 0)$&$s_0$ flip&$(1, 0, 0, 0, 0)$\\
$YIIII$&$(1, 0, 1, 1, 1)$&$s_1$ flip&$(0, 1, 0, 0, 0)$\\
$IYIII$&$(1, 1, 0, 1, 1)$&$s_2$ flip&$(0, 0, 1, 0, 0)$\\
$IIYII$&$(1, 1, 1, 0, 1)$&$s_3$ flip&$(0, 0, 0, 1, 0)$\\
$IIIYI$&$(1, 1, 1, 1, 0)$&$s_4$ flip&$(0, 0, 0, 0, 1)$\\
$IIIIY$&$(0, 1, 1, 1, 1)$&&
\end{tabular}
\end{ruledtabular}
\end{table}
In fact, the same technique works for any single-error-correcting stabilizer code.
\begin{theorem}\label{th:single}
For any $[[n,k,3]]$ stabilizer code, there exists a set of at most $n-k+1$ stabilizer operators
that distinguish all single errors and no error among data qubits and syndrome bits that have distinct effects on the encoded quantum information.
\end{theorem}
\Proof
Let $\mathcal{G}$ be a set of $n-k$ independent generators of the stabilizer of an $[[n,k,3]]$ stabilizer code.
Define $G' = \prod_{G\in\mathcal{G}}G$ as the product of $n-k$ generators in $\mathcal{G}$.
Let $\boldsymbol{s}_E, \boldsymbol{s}'_E$ be the syndromes of an error $E$ on data qubits given by $\mathcal{G}$ only and by $\mathcal{G}\cup\{G'\}$ respectively.
Because $\mathcal{G}$ generates the stabilizer of an $[[n,k,3]]$ stabilizer code,
it is trivial that for any pair $E_0, E_1$ of single errors that have different effects on the encoded quantum information,
we have $\boldsymbol{s}'_{E_0} \not= \boldsymbol{s}'_{E_1}$.
Because $G'$ is the product of generators in $\mathcal{G}$, the extra syndrome bit by $G'$ is $0$ if $\operatorname{wt}(\boldsymbol{s}_E)$ is even and $1$ otherwise.
Hence, we have
\[
\operatorname{wt}(\boldsymbol{s}'_E) =
\begin{cases}
\operatorname{wt}(\boldsymbol{s}_E) &\mbox{if } \operatorname{wt}(\boldsymbol{s}_E) \mbox{ is even}\\
\operatorname{wt}(\boldsymbol{s}_E)+1 &\mbox{otherwise}
\end{cases},
\]
which implies that $\operatorname{wt}(\boldsymbol{s}'_E) \not= 1$ when there is an erroneous data qubit.
Because all single errors on syndrome bits result in syndromes of weight $1$,
if the syndrome bit by the redundant stabilizer operator $G'$ is extracted along with the other $n-k$ syndrome bits,
single errors on syndrome bits result in different syndromes from any correctable error on data qubits.
When a single error occurs on the extracted syndrome, the erroneous syndrome bit is identified as the one whose value is $1$.
\qed

More curious, perhaps, is that redundant stabilizer operators are not always necessary.
For instance, the \textit{Steane code} \cite{Steane:1996a} is typically presented
as a $[[7,1,3]]$ \textit{Calderbank-Shor-Steane} (CSS) \textit{code} \cite{Calderbank:1996,Steane:1996} with generators
\begin{align*}
S_0 &= XIIXIXX, \ S_1 = IXIXXIX, \ S_2 = IIXIXXX,\\
S_3 &= ZIIZIZZ, \ \ \ S_4 = IZIZZIZ, \ \ \ S_5 = IIZIZZZ.
\end{align*}
At first blush, it may appear that this code also needs one more stabilizer operator to become globally single-error-correcting.
In fact, the correct syndrome of $Z$ acting on the first qubit is $(1,0,0,0,0,0)$,
which is indistinguishable from a plain bit flip on $s_0$.
However, this is due to the choice of generators.
The following independent generators of the Steane code distinguish all single errors on data qubits and syndrome bits
\begin{align*}
S'_0 &= S_0S_3, & S'_1 &= S_1S_3, & S'_2 &= S_2S_3,\\
S'_3 &= S_3\prod_{i=0}^5 S_i, & S'_4 &= S_4\prod_{i=0}^5 S_i, & S'_5 &= S_5\prod_{i=0}^5 S_i.
\end{align*}
The alternative six independent generators $S'_i$ can be written as
\begin{align*}
\left[
\begin{array}{c}
S'_0\\
S'_1\\
S'_2\\
S'_3\\
S'_4\\
S'_5
\end{array}
\right]
=
\left[
\begin{array}{ccccccc}
Y&I&I&Y&I&Y&Y\\
Z&X&I&Y&X&Z&Y\\
Z&I&X&Z&X&Y&Y\\
X&Y&Y&Z&I&Z&X\\
Y&X&Y&Z&Z&I&X\\
Y&Y&X&I&Z&Z&X\\
\end{array}
\right].
\end{align*}
Table \ref{tbl:steane} lists the syndrome of each single error
by the original generators $S_i$ of CSS type and the alternative minimal generating set.
\begin{table}[h!t]\caption{Syndromes by the Steane code.\label{tbl:steane}}
\begin{ruledtabular}
\begin{tabular}{ccc}
Error&$(s_0, s_1, s_2, s_3,s_4,s_5)$&$(s'_0, s'_1, s'_2, s'_3,s'_4,s'_5)$\\\hline
No error&$(0, 0, 0, 0, 0, 0)$&$(0, 0, 0, 0, 0, 0)$\\
$XIIIIII$&$(0, 0, 0, 1, 0, 0)$&$(1, 1, 1, 0, 1, 1)$\\
$IXIIIII$&$(0, 0, 0, 0, 1, 0)$&$(0, 0, 0, 1, 0, 1)$\\
$IIXIIII$&$(0, 0, 0, 0, 0, 1)$&$(0, 0, 0, 1, 1, 0)$\\
$IIIXIII$&$(0, 0, 0, 1, 1, 0)$&$(1, 1, 1, 1, 1, 0)$\\
$IIIIXII$&$(0, 0, 0, 0, 1, 1)$&$(0, 0, 0, 0, 1, 1)$\\
$IIIIIXI$&$(0, 0, 0, 1, 0, 1)$&$(1, 1, 1, 1, 0, 1)$\\
$IIIIIIX$&$(0, 0, 0, 1, 1, 1)$&$(1, 1, 1, 0, 0, 0)$\\
$YIIIIII$&$(1, 0, 0, 1, 0, 0)$&$(0, 1, 1, 1, 0, 0)$\\
$IYIIIII$&$(0, 1, 0, 0, 1, 0)$&$(0, 1, 0, 0, 1, 0)$\\
$IIYIIII$&$(0, 0, 1, 0, 0, 1)$&$(0, 0, 1, 0, 0, 1)$\\
$IIIYIII$&$(1, 1, 0, 1, 1, 0)$&$(0, 0, 1, 1, 1, 0)$\\
$IIIIYII$&$(0, 1, 1, 0, 1, 1)$&$(0, 1, 1, 0, 1, 1)$\\
$IIIIIYI$&$(1, 0, 1, 1, 0, 1)$&$(0, 1, 0, 1, 0, 1)$\\
$IIIIIIY$&$(1, 1, 1, 1, 1, 1)$&$(0, 0, 0, 1, 1, 1)$\\
$ZIIIIII$&$(1, 0, 0, 0, 0, 0)$&$(1, 0, 0, 1, 1, 1)$\\
$IZIIIII$&$(0, 1, 0, 0, 0, 0)$&$(0, 1, 0, 1, 1, 1)$\\
$IIZIIII$&$(0, 0, 1, 0, 0, 0)$&$(0, 0, 1, 1, 1, 1)$\\
$IIIZIII$&$(1, 1, 0, 0, 0, 0)$&$(1, 1, 0, 0, 0, 0)$\\
$IIIIZII$&$(0, 1, 1, 0, 0, 0)$&$(0, 1, 1, 0, 0, 0)$\\
$IIIIIZI$&$(1, 0, 1, 0, 0, 0)$&$(1, 0, 1, 0, 0, 0)$\\
$IIIIIIZ$&$(1, 1, 1, 0, 0, 0)$&$(1, 1, 1, 1, 1, 1)$\\
$s_0$ flip&$(1, 0, 0, 0, 0, 0)$&N/A\\
$s_1$ flip&$(0, 1, 0, 0, 0, 0)$&N/A\\
$s_2$ flip&$(0, 0, 1, 0, 0, 0)$&N/A\\
$s_3$ flip&$(0, 0, 0, 1, 0, 0)$&N/A\\
$s_4$ flip&$(0, 0, 0, 0, 1, 0)$&N/A\\
$s_5$ flip&$(0, 0, 0, 0, 0, 1)$&N/A\\
$s'_0$ flip&N/A&$(1, 0, 0, 0, 0, 0)$\\
$s'_1$ flip&N/A&$(0, 1, 0, 0, 0, 0)$\\
$s'_2$ flip&N/A&$(0, 0, 1, 0, 0, 0)$\\
$s'_3$ flip&N/A&$(0, 0, 0, 1, 0, 0)$\\
$s'_4$ flip&N/A&$(0, 0, 0, 0, 1, 0)$\\
$s'_5$ flip&N/A&$(0, 0, 0, 0, 0, 1)$
\end{tabular}
\end{ruledtabular}
\end{table}

Note that if we would like to maintain the CSS property that each stabilizer operator is composed of $I$ and $X$ only or $I$ and $Z$ only,
we need $2$ extra stabilizer operators.
For this purpose, the stabilizer operators $\prod_{i=0}^{2}S_i$ and $\prod_{i=3}^{5}S_i$ work.
Because the classical linear code underlying the Steane code is a perfect code,
this is an unavoidable penalty for being globally single-error-correcting and maintaining the CSS property.
In general, global single error correction can be achieved while maintaing the CSS property
by adding a pair of stabilizer operators analogously to Theorem \ref{th:single} if there is no good choice of independent generators.

\subsection{Global double error correction}\label{subsec:double}
More attractive may be double-error-correcting codes because they can offer stronger protection against decoherence.
The concept of \textit{perfect hash families} \cite{Mehlhorn:1984} assures that the cost of extending double error correction is at most logarithmic,
even if double errors include two incorrect syndrome bits as well as one data qubit and one syndrome bit being simultaneously erroneous.
\begin{theorem}\label{th:DoubleMain}
For any $[[n,k,5]]$ stabilizer code, there exists a collection of at most $n-k+2\lceil\log_2(n-k)\rceil+3$ stabilizer operators
that distinguish all single, double, and no errors among data qubits and syndromes bits
that have distinct effects on the encoded quantum information.
\end{theorem}

To verify Theorem \ref{th:DoubleMain}, we first prove a lemma, which uses a binary vector to represent an operator on qubits.
For an $n$-fold tensor product $P=O_0,\otimes\dots\otimes O_{n-1}$ of operators $O_i \in \{I,X,Y,Z\}$,
the \textit{error vector} of $P$ is the $2n$-dimensional vector $\boldsymbol{v} = (v_0,\dots,v_{2n-1}) \in \mathbb{F}_2^{2n}$
over the finite field $\mathbb{F}_2$ of order $2$ such that for $0\leq i \leq n-1$
\[
v_i =\begin{cases}
0 &\text{if } O_i = I, Z,\\
1 &\text{otherwise}
\end{cases}
\]
and
\[
v_{i+n} =\begin{cases}
0 &\text{if } O_i = I, X,\\
1 &\text{otherwise}.
\end{cases}
\]
Ignoring the overall factor $i^\lambda$, we may speak of the error vector of any $P \in \mathcal{P}$ including stabilizer operators of a stabilizer code.
Given a set $\mathcal{O}$ of $m$ stabilizer operators of an $[[n,k,d]]$ stabilizer code,
a \textit{quantum parity-check matrix} specified by $\mathcal{O}$ is an $m\times 2n$ binary matrix
whose rows are the error vectors of stabilizer operators in $\mathcal{O}$.
\begin{lemma}\label{lm:ForDouble}
Let $H$ be an $(n-k+r) \times 2n$ quantum parity-check matrix of an $[[n,k,d]]$ stabilizer code
specified by a set of $n-k$ independent generators and $r$ redundant stabilizer operators.
The corresponding $n-k+r$ stabilizer operators produce different syndromes for all patterns
of errors on up to $\left\lfloor\frac{d-1}{2}\right\rfloor$ data qubits and/or syndromes bits
that have different effects from each other on the encoded quantum information if
any error vector $\boldsymbol{e} \in \mathbb{F}_2^{2n}$ corresponding to an error on $t$ qubits with $t \leq d-1$ satisfies that
$\operatorname{wt}\mkern-\medmuskip\left(H\boldsymbol{e}^T\right) \geq d-t$ or that $H\boldsymbol{e}^T = \boldsymbol{0}$.
\end{lemma}
\Proof
We consider a slightly stronger condition that any pair of errors, one of which is on up to $\left\lfloor\frac{d-1}{2}\right\rfloor$ data qubits and/or syndromes bits
and the other of which is on up to $\left\lfloor\frac{d}{2}\right\rfloor$ data qubits and/or syndromes bits, give
different syndromes if they have different effects from each other on the encoded quantum information.
Let $t_0$, $t_1$ be a pair of positive integers such that $t_0 \leq \left\lfloor\frac{d}{2}\right\rfloor$ and  $t_1 \leq \left\lfloor\frac{d-1}{2}\right\rfloor$.
Take arbitrary error vectors $\boldsymbol{e}_0$ and $\boldsymbol{e}_1$ corresponding to errors of weight $t_0$ and $t_1$ respectively.
Assume that there may be errors on up to $\left\lfloor\frac{d}{2}\right\rfloor-t_0$ and $\left\lfloor\frac{d-1}{2}\right\rfloor-t_1$ syndrome bits
when extracting the syndromes of $\boldsymbol{e}_0$ and $\boldsymbol{e}_1$ respectively.
We let $(n-k+r)$-dimensional binary vectors
$\boldsymbol{f}_0 = (f^{(0)}_0,\dots,f^{(0)}_{n-1}), \boldsymbol{f}_1 = (f^{(1)}_0,\dots,f^{(1)}_{n-1}) \in \mathbb{F}_2^{n-k+r}$ represent the errors on syndromes
by defining $f^{(i)}_j = 1$ if the $j$th syndrome bit is flipped when extracting the syndrome of $\boldsymbol{e}_i$ and $0$ otherwise.
By assumption, we have $\operatorname{wt}(\boldsymbol{f}_0) \leq \left\lfloor\frac{d}{2}\right\rfloor-t_0$
and $\operatorname{wt}(\boldsymbol{f}_1) \leq \left\lfloor\frac{d-1}{2}\right\rfloor-t_1$.
The two errors give the same syndrome if and only if
\[H\boldsymbol{e}_0^T + \boldsymbol{f}_0^T = H\boldsymbol{e}_1^T + \boldsymbol{f}_1^T,\]
which holds if and only if
\[H(\boldsymbol{e}_0+\boldsymbol{e}_1)^T = (\boldsymbol{f}_0+\boldsymbol{f}_1)^T.\]
Note that the errors corresponding to $\boldsymbol{e}_0$ and $\boldsymbol{e}_1$ have the same effect on the encoded quantum information
if and only if the $n$-fold tensor product of Pauli operators that correspond to $\boldsymbol{e}_0+\boldsymbol{e}_1$ is a stabilizer operator.
Because $t_0+t_1< d$, this is equivalent to the condition that $H(\boldsymbol{e}_0+\boldsymbol{e}_1)^T = 0$.
Note also that
\begin{align*}
\operatorname{wt}(\boldsymbol{f}_0+\boldsymbol{f}_1) &\leq \left\lfloor\frac{d}{2}\right\rfloor - t_0 + \left\lfloor\frac{d-1}{2}\right\rfloor - t_1\\
&=d-t_0-t_1-1.
\end{align*}
Thus, by rewriting $\boldsymbol{e}_0+\boldsymbol{e}_1$ and $t_0+t_1$ as $\boldsymbol{e}$ and $t$ respectively,
the $n-k+r$ stabilizer operators produce different syndromes for all patterns
of up to $\left\lfloor\frac{d-1}{2}\right\rfloor$ errors among data qubits and syndromes bits
that have different effects from each other on the encoded quantum information
if any error vector $\boldsymbol{e} \in \mathbb{F}_2^{2n}$ corresponding to an error of weight $t \leq d-1$ satisfies that
$\operatorname{wt}\mkern-\medmuskip\left(H\boldsymbol{e}^T\right) \geq d-t$ or that $H\boldsymbol{e}^T = \boldsymbol{0}$ as desired.
\qed

To prove Theorem \ref{th:DoubleMain}, we use a special set of functions.
A $(w,v)$-\textit{hash function} is a function $h : A \rightarrow B$ between finite sets $A$ and $B$, where $\vert A \vert = w$ and $\vert B \vert = v$.
The function $h$ is \textit{perfect} with respect to a subset $X \subseteq A$ if $h$ is injective on $X$, that is, if $h\vert_X$ is one-to-one.
Let $F$ be a set of $m$ $(w, v)$-hash functions between $A$ and $B$, where $w \geq v \geq t \geq 2$.
Then $F$ is a \textit{perfect hash family} \textup{PHF}$(m; w, v, t)$
if for any $X \subseteq A$ with $\vert X \vert = t$, there exists at least one $h \in F$ such that $h\vert_X$ is one-to-one.

We employ a perfect hash family with $v=t=2$. In this case, there is a convenient representation in terms of binary matrix.
A perfect hash family PHF$(m; w, 2, 2)$ is equivalent to an $m \times w$ matrix over $\mathbb{F}_2$ in which
any pair of columns has at least one row whose entries sum to $1$.
This is equivalent to say that any $m\times 2$ submatrix has $(0,1)$ or $(1,0)$ somewhere in their rows.
The equivalence can be seen straightforwardly
by indexing rows and columns of $M$ by functions in $F$ and elements of $A$ respectively,
so that the entry of column $i$ of the row $h$ represents the value of $h(i)$.

A PHF$(m; 2^m, 2, 2)$ can be constructed by taking all distinct $m$-dimensional binary columns.
Deleting a column from a perfect hash family gives another one with fewer columns.
Hence, a PHF$(m,w,2,2)$ exists for $m = \lceil\log_2w\rceil$.

$ $\\
\noindent\textbf{Proof of Theorem \ref{th:DoubleMain}.}\quad
Let $H$ be an $(n-k) \times 2n$ quantum parity-check matrix of an $[[n,k,5]]$ stabilizer code.
Let $m=\lceil\log_2(n-k)\rceil$.
We define $2m+3$ redundant stabilizer operators to be joined.
Write the $i$th row of $H$ as $\boldsymbol{h}^{(i)}$.
Let $M$ be an $m \times (n-k)$ binary matrix forming a PHF$(m;n-k,2,2)$.
Write the $i$th row of $M$ as $\boldsymbol{r}^{(i)} = (r_0^{(i)},\dots,r_{n-k-1}^{(i)})$.
Let $N$ be the $m \times 2n$ binary matrix $N$ whose $i$th row $\boldsymbol{n}^{(i)}$ is defined by
\begin{align}\label{def:N}
\boldsymbol{n}^{(i)} = \sum_{j \in \{l \mid r_l^{(i)} = 1\}}\boldsymbol{h}^{(j)},
\end{align}
where addition is over $\mathbb{F}_2^{2n}$.
Let $A$ be the $3 \times 2n$ binary matrix in which each row is the sum of the $n-k$ rows in $H$ over $\mathbb{F}_2^{2n}$.
Note that the rows of $H$, $N$, and $A$ all correspond to stabilizer operators of the $[[n,k,5]]$ stabilizer code.
Let $S$ be the $(n-k+2m+3) \times 2n$ quantum parity-check matrix defined by $n-k+2m+3$ stabilizer operators as follows:
\[S = \left[\begin{array}{c}H\\A\\N\\N\\\end{array}\right].\]

We show that $S$ gives different syndromes for all patterns of up to two errors among data qubits and syndromes bits
that have different effects from each other on encoded quantum information.
By Lemma \ref{lm:ForDouble}, we only need to check whether
any error vector $\boldsymbol{e} \in \mathbb{F}_2^{2n}$ corresponding to an error of weight $t \leq 4$ which is not a stabilizer operator satisfies the condition that
$\operatorname{wt}\mkern-\medmuskip\left(S\boldsymbol{e}^T\right) \geq 5-t$.

Let $W$ be the set of coordinates $i$ such that $e_i = 1$, where $\boldsymbol{e} = (e_0,\dots,e_{2n-1})$.
Note that because any $[[n,k,5]]$ stabilizer code obeys the quantum Singleton bound $n-k \geq 2(d-1)$,
the condition that $t \leq 4$ implies that $\vert W \vert = \operatorname{wt}(\boldsymbol{e}) \leq 2t \leq n-k$.
We write the $i$th columns of $S$, $H$, $A$, and $N$ as $\boldsymbol{s}^{(i)}$, $\boldsymbol{c}^{(i)}$, $\boldsymbol{a}^{(i)}$, and $\boldsymbol{p}^{(i)}$ respectively.
If $S\boldsymbol{e}^T = \boldsymbol{0}$, it is a harmless error.
We assume that $\boldsymbol{e}$ corresponds to a harmful error that acts nontrivially on the encoded quantum information.
Thus, we have
\begin{align}\label{Cpositive}
\operatorname{wt}\mkern-\medmuskip\left(H\boldsymbol{e}^T\right) &= \operatorname{wt}\mkern-\medmuskip\left(\sum_{i \in W}\boldsymbol{c}^{(i)}\right)\notag\\
&> 0.
\end{align}
First we consider the case $\operatorname{wt}\mkern-\medmuskip\left(\sum_{i\in W}\boldsymbol{a}^{(i)}\right)=0$.
Because $\operatorname{wt}\mkern-\medmuskip\left(\sum_{i\in W}\boldsymbol{a}^{(i)}\right)=0$ if and only if
$\operatorname{wt}\mkern-\medmuskip\left(\sum_{i\in W}\boldsymbol{c}^{(i)}\right)$ is even, we have
\[\operatorname{wt}\mkern-\medmuskip\left(\sum_{i \in W}\boldsymbol{c}^{(i)}\right) \geq 2,\]
where the left-hand side is even. If
\[\operatorname{wt}\mkern-\medmuskip\left(\sum_{i \in W}\boldsymbol{c}^{(i)}\right) \geq 4,\]
then $\operatorname{wt}\mkern-\medmuskip\left(S\boldsymbol{e}^T\right) \geq 4$ as desired.
Hence, we only need to consider the situation where there exist exactly two coordinates at which the entries of $\sum_{i \in W}\boldsymbol{c}^{(i)}$ are $1$.
Let $a$ and $b$ be these two coordinates.
By the definition of a perfect hash family, there exists at least one row $\boldsymbol{r}^{(j)} = (r_0^{(j)},\dots,r_{n-k-1}^{(j)})$ in $M$ such that $r_a^{(j)}+r_b^{(j)}=1$.
Hence, by Equation (\ref{def:N}), $N\boldsymbol{e}^T$ contains a row which is the same as either $h^{(a)}\boldsymbol{e}^T$ or $h^{(b)}\boldsymbol{e}^T$,
either of which is $1$.
Thus, we have
\[\operatorname{wt}\mkern-\medmuskip\left(\sum_{i \in W}\boldsymbol{p}^{(i)}\right) \geq 1.\]
Because we have two copies of $N$ in $S$, we have
\begin{align*}
\operatorname{wt}\mkern-\medmuskip\left(S\boldsymbol{e}^T\right) &= \operatorname{wt}\mkern-\medmuskip\left(\sum_{i \in W}\boldsymbol{s}^{(i)}\right)\\
&=\operatorname{wt}\mkern-\medmuskip\left(\sum_{i \in W}\boldsymbol{c}^{(i)}\right)
+\operatorname{wt}\mkern-\medmuskip\left(\sum_{i \in W}\boldsymbol{a}^{(i)}\right)\\
&\quad+2\operatorname{wt}\mkern-\medmuskip\left(\sum_{i \in W}\boldsymbol{p}^{(i)}\right)\\
&\geq 2+0+2\\
&= 4.
\end{align*}
Thus, for any positive integer $t$, we have $\operatorname{wt}\mkern-\medmuskip\left(S\boldsymbol{e}^T\right)\geq 5-t$.
The remaining case is when $\operatorname{wt}\mkern-\medmuskip\left(\sum_{i\in W}\boldsymbol{a}^{(i)}\right)\not=0$.
Because each row of $A$ is the sum of the $n-k$ rows of $H$, this means that $\operatorname{wt}\mkern-\medmuskip\left(\sum_{i\in W}\boldsymbol{a}^{(i)}\right)=3$.
By Inequality (\ref{Cpositive}), we have
\begin{align*}
\operatorname{wt}\mkern-\medmuskip\left(S\boldsymbol{e}^T\right) &\geq
\operatorname{wt}\mkern-\medmuskip\left(\sum_{i \in W}\boldsymbol{c}^{(i)}\right)
+\operatorname{wt}\mkern-\medmuskip\left(\sum_{i \in W}\boldsymbol{a}^{(i)}\right)\\
&\geq 1+3\\
&= 4.
\end{align*}
The proof is complete.
\qed

\subsection{Asymmetric global error correction}
In the previous two sections, we showed how to make stabilizers globally single- and double-error-correcting
without changing the Hilbert spaces they stabilize.
In principle, we could consider global $t$-error correction for data qubits and syndrome bits for $t \geq 3$ as well.
However, such an approach would be suboptimal if the error probability of data qubits is different from that of syndrome bits, which is very likely the case in practice.
Hence, in a situation where more powerful error correction than single or double error correction is required,
it is more reasonable to treat error correction for data qubits and syndrome bits separately.
In this section, we study a set of stabilizer operators of an $[[n,k,d]]$ stabilizer code
that is $\lfloor(d-1)/2\rfloor$-error-correcting for data qubits and $\lfloor(t-1)/2\rfloor$-error-correcting for syndrome bits for given $t$. 

To study such asymmetric global error correction, we introduce a useful view of what a whole stabilizer looks like.
Take a set $Q$ of $n$ qubits.
The $l$-\textit{local action} of $P\in\mathcal{P}$ on a subset $L\subseteq Q$ with $\vert L\vert=l$ is the $l$-fold tensor product
obtained by discarding the overall factor $i^\lambda$ and operators acting on the $n-l$ qubits not in $L$.
Delsarte's equivalence theorem \cite{Delsarte:1973} in algebraic combinatorics shows that stabilizer codes are everywhere locally completely stochastic.
\begin{theorem}\label{thm:Delsarte}
Let $\mathcal{S}$ be the stabilizer of a stabilizer code of pure distance $d_p$ and $L$ a set of $l$ data qubits with $l<d_p$.
Take uniformly at random a stabilizer operator $S\in\mathcal{S}$ and let $A_L$ be its $l$-local action on $L$.
For any $l$-fold tensor product $T$ of operators $O_i\in\{I,X,Y,Z\}$,
the probability that $A_L=T$ is $4^{-l}$.
\end{theorem}

To prove the above theorem, we first give a proposition.
We write the finite field of order $q^r$ with $q$ prime power as $\mathbb{F}_{q^r}$.
An \textit{inner product} over the elementary abelian group $\mathbb{Z}_v^n$ of order $v^n$ is a symmetric biadditive form $B$
such that $B(\boldsymbol{a}, \boldsymbol{b}) = B(\boldsymbol{a}, \boldsymbol{c})$ holds for any $\boldsymbol{a}\in\mathbb{Z}_v^n$
if and only if $\boldsymbol{b}=\boldsymbol{c}$.
An $\mathbb{F}_q$-\textit{additive code} $\mathcal{C}$ of \textit{length} $n$, \textit{dimension} $k$, and \textit{minimum distance} $d$ over $\mathbb{F}_{q^r}$
is an additive subgroup of $\mathbb{F}_{q^r}^n$ of order $\vert\mathcal{C}\vert$ such that
$\log_q(\vert\mathcal{C}\vert)=k$ and $\min\{\operatorname{wt}(\boldsymbol{c})\mid\boldsymbol{c}\in\mathcal{C}\setminus\{\boldsymbol{0}\}\}=d$.
Each element of $\mathcal{C}$ is a \textit{codeword}.
The \textit{dual} of $\mathcal{C}$ with respect to inner product $B$ is the additive code
$\mathcal{C}^{\perp} = \{\boldsymbol{c}'\mid B(\boldsymbol{c},\boldsymbol{c}')=\boldsymbol{0} \text{\ for any\ } \boldsymbol{c}\in\mathcal{C}\}$.
The \textit{dual distance} $d^{\perp}$ of $\mathcal{C}$ is the minimum distance of $\mathcal{C}^{\perp}$.
An \textit{orthogonal array} $\textup{OA}(u,n,v,s)$ is an $u\times n$ matrix over a finite set $\Gamma$ of cardinarity $v$ such that
in any $u\times s$ submatrix every $s$-dimensional vector in $\Gamma^s$ appears exactly $\frac{u}{v^s}$ times as a row.
The following is a straightforward corollary of Delsarte's equivalence theorem \cite[Theorem 4.5]{Delsarte:1973} in algebraic combinatorics.
\begin{proposition}\label{prop:linearDelsarte}
Let $\mathcal{C}$ be an $\mathbb{F}_q$-additive code  over $\mathbb{F}_{q^r}$
of length $n$, dimension $k$, and dual distance $d^{\perp}$ with respect to some inner product $B$.
A $q^k\times n$ matrix formed by all codewords of $C$ as rows is an $\textup{OA}(q^k,n,q^r,d^{\perp}-1)$.
\end{proposition}

Now we are ready to prove Theorem \ref{thm:Delsarte}.

\noindent\textbf{Proof of Theorem \ref{thm:Delsarte}.}\quad
Let $\mathcal{S}$ be the stabilizer of an $[[n,k]]$ stabilizer code whose pure distance is $d_p$.
For each stabilizer operator $S=i^\lambda O_0\otimes\dots\otimes O_{n-1}\in\mathcal{S}$, define its corresponding $n$-dimensional vector
$\boldsymbol{c}^{(S)} = (c_0^{(S)},\dots,c_{n-1}^{(S)}) \in \mathbb{F}_4^{n}$ over the finite field $\mathbb{F}_4 = \{0,1,\omega,\omega^2=\omega+1\}$ of order $4$
such that
\[c_i^{(S)}=
\begin{cases}
0 &\mbox{if } O_i = I,\\
1 &\mbox{if } O_i = Y,\\
\omega &\mbox{if } O_i = X,\\
\omega^2 &\mbox{if } O_i = Z.
\end{cases}\]
The set $\mathcal{C} = \{\boldsymbol{c}^{(S)} \mid S\in\mathcal{S}\}$ is an $\mathbb{F}_2$-additive code of length $n$, dimension $n-k$, and dual distance $d_p$
over $\mathbb{F}_4$ (see \cite{Calderbank:1998}).
Thus, by Proposition \ref{prop:linearDelsarte}, a $2^{n-k}\times n$ matrix $M$ formed by all codewords of $C$ as rows is an $\textup{OA}(2^{n-k},n,4,d_p-1)$.
By definition an $\textup{OA}(2^{n-k},n,4,d_p-1)$ is an $\textup{OA}(2^{n-k},n,4,l)$ for any $l \leq d_p-1$ as well.
Thus, in any $2^{n-k}\times l$ submatrix of $M$, every $l$-dimensional vector in $\mathbb{F}_4^l$ appears exactly $2^{n-k-2l}$ times as a row.
Hence, given an $l$-dimensional vector $\boldsymbol{v}\in\mathbb{F}_4^l$ and $2^{n-k}\times l$ submatrix of $M$,
the probability that a uniformly randomly chosen row is $\boldsymbol{v}$ is $2^{n-k-2l-(n-k)}=4^{-l}$.
\qed

We consider how many stabilizer operators are sufficient to correct $\lfloor(t-1)/2\rfloor$ erroneous syndrome bits for a given positive integer $t$.
Because the use of redundant stabilizer operators changes the number of syndrome bits we needs to take care of,
it is natural to aim for correcting all errors of weight up to a fixed fraction of the total number $m$ of extracted syndrome bits rather than an absolute constant.
Hence, we let $t=\lceil\delta m\rceil$ for some positive constant $\delta<1/2$.

A fundamental lower bound on the achievable rate $k/n$ for an $[[n,k,d]]$ stabilizer code is the \textit{quantum Gilbert-Varshamov bound} \cite{Ekert:1996}.
It states that for any positive integers $n$, $k$, and $d$ such that
\[\sum_{i=1}^{d-1}3^i\binom{n}{i}\leq2^{n-k},\]
there exists an $[[n,k,d]]$ stabilizer code.
In fact, it can be shown that the stabilizer code can be nondegenerate so that $d=d_p$.
We consider how many stabilizer operators are necessary for a nondegenerate stabilizer code meeting this bound to overcome a reasonable number of syndrome bit errors.

In what follows, $H_2(x) = -x\log_2x-(1-x)\log_2(1-x)$ is the binary entropy function.
We use probabilistic combinatorics \cite{Alon:2008} to exploit the local randomness of stabilizer codes.
\begin{theorem}\label{th:cssMain}
Let $\mathcal{S}$ be the stabilizer of an $[[n,k,d]]$ stabilizer code of pure distance $d_p=d$ that obeys the quantum Gilbert-Varshamov bound.
Take a constant $\delta$ such that $0<\delta<1/2$ and define
\[m = \left\lceil\frac{n-k}{1-H_2(\delta)}\right\rceil.\]
There exists a collection $\mathcal{C}$ of at most $m$ stabilizer operators chosen from $\mathcal{S}$
that corrects an arbitrary error on up to $\lfloor(d_p-1)/2\rfloor$ data qubits and up to $\lfloor(t-1)/2\rfloor$ syndrome bits,
where $t=\lceil\delta m\rceil$.
\end{theorem}
\Proof
If $\mathcal{C}$ never gives the all-zero syndrome when up to $d_p-1$ data qubits and up to $t-1$ syndrome bits are erroneous except when there is no error,
then $\mathcal{C}$ corrects an arbitrary error on up to $\lfloor(d_p-1)/2\rfloor$ data qubits and up to $\lfloor(t-1)/2\rfloor$ erroneous syndrome bits.
Indeed, the condition assures that all patterns of up to $\lfloor(d_p-1)/2\rfloor$ erroneous data qubits and up to $\lfloor(t-1)/2\rfloor$ incorrect syndrome bits
result in distinct syndromes.

Pick uniformly at random $m$ stabilizer operators in $\mathcal{S}$ allowing repetition.
Take a nontrivial error $E\in\mathcal{P}$ of weight $l$ such that $1 \leq l\leq d_p-1$.
By Theorem \ref{thm:Delsarte},
the probability that the $m$ syndrome bits can be all $0$ for $E$ when up to $t-1$ syndrome bits are flipped is
\[p_E=2^{-m}\sum_{i=0}^{t-1}\binom{m}{i}.\]
Let $V$ be the random variable counting the number of nontrivial errors of weight less than $d_p$ that result in the all-zero syndrome due to up to $t-1$ syndrome bit flips.
Its expected value is
\[\mathbb{E}[V]=2^{-m}\sum_{i=0}^{t-1}\binom{m}{i}\sum_{j=1}^{d_p-1}3^j\binom{n}{j}.\]
Note that because our nondegenerate stabilizer code obeys the quantum Gilbert-Varshamov bound, we have
\begin{align*}\label{ieq:hamming}
\sum_{j=0}^{d_p-1}3^j\binom{n}{j}\leq2^{n-k}.
\end{align*}
Hence, by applying the following bound on the partial sum of binomial coefficients
\[\sum_{i=0}^{\lfloor\delta m\rfloor}\binom{m}{i}\leq2^{mH_2(\delta)}\]
(see, for example, \cite{MacWilliams:1977}),
we have
\[\mathbb{E}[V]<2^{m(H_2(\delta)-1)+n-k}.\]
If $\mathbb{E}[V]<1$, there exists a collection of $m$ stabilizer operators
in which no combination of a nontrivial error of weight less than or equal to $d_p-1$ and up to $t-1$ syndrome bit flips results in the all-zero syndrome.
$\mathbb{E}[V]<1$ holds if
\[m\geq\frac{n-k}{1-H_2(\delta)}.\]
Noting that $m$ must be an integer, the above inequality shows that
\[m=\left\lceil\frac{n-k}{1-H_2(\delta)}\right\rceil\]
is sufficient as desired.
\qed

\subsection{Hybrid Hamming bound}
In this subsection, we turn our attention to how many redundant stabilizer operators are necessary instead of how many are sufficient.
The \textit{Hamming bound} \cite{MacWilliams:1977} describes a fundamental limit on the parameters of a classical error-correcting code.
There is a quantum analogue, called the \textit{quantum Hamming bound} \cite{Ekert:1996,Gottesman:1996}.
By counting the combinations of quantum errors and classical bit flips,
we obtain a hybrid Hamming bound for a scheme that protects a physical system holding both quantum and classical information.
\begin{theorem}\label{th:qcHamming}
Take $n_q$ qubits and $n_c$ bits.
If $s$-bit classical information distinguishes
all combinations of discretized errors on up to $t_q$ qubits and up to $t_c$ bits, then
\[\sum_{i=0}^{t_q}\sum_{j=0}^{t_c}3^i\binom{n_q}{i}\binom{n_c}{j} \leq 2^{s}.\]
\end{theorem}
This reduces to the classical Hamming bound for codes decodable by syndromes, such as linear codes,
by setting $n_q=0$ and the quantum Hamming bound by setting $n_c=0$.
Assuming an $[[n,k,d]]$ stabilizer code with $r$ redundant stabilizer operators,
plugging $n_q=n$, $n_c=s=n-k+r$ gives 
\[\sum_{i=0}^{t_q}\sum_{j=0}^{t_c}3^i\binom{n}{i}\binom{n-k+r}{j} \leq 2^{n-k+r}.\]

For symmetric global $t$-error correction that uses one same distance parameter for both quantum errors and classical bit flips
as in Sections \ref{subsec:single} and \ref{subsec:double}, we have
\[\sum_{i=0}^{t-j}\sum_{j=0}^{t}3^i\binom{n}{i}\binom{n-k+r}{j} \leq 2^{n-k+r}.\]

It should be noted that, as in the standard quantum Hamming bound, the hybrid bounds only apply to schemes that do not exploit degeneracy.
As we have seen in the proofs of Theorems \ref{th:single} and \ref{th:DoubleMain},
stabilizer codes can take advantage of degeneracy when correcting combinations of erroneous data qubits and flipped syndrome bits.
While no stabilizer codes are known to violate the quantum Hamming bound,
more efficient stabilizer codes are not entirely ruled out.

\section{Relation to fault-tolerant syndrome extraction}\label{sec:ft}
In this section we relate stabilizer codes' ability to correct imperfect syndromes to Shor's syndrome extraction for fault-tolerant quantum computation.
A concise introduction to fault-tolerant quantum computation can be found in \cite{Gottesman:2010a}.

Assuming each quantum gate is implemented with standard fault-tolerant techniques,
Shor's method extract the information about the eigenvalue of an error for each stabilizer operator in succession.
Abstractly, this means that syndrome bits are obtained one-by-one in a sequence.
During this process, any physical qubit including one for storing a syndrome bit can spontaneously decohere.
Each quantum gate involved in extracting a syndrome bit may also introduce errors on qubits it interacts with.

In general, we would like to know what error there was when syndrome extraction started and what error has been introduced since then.
More precisely, our task is to infer a most likely \textit{fault path} that is consistent with the extracted syndrome under a given error model
(see, for example, \cite{Gottesman:2014}).

Such inference needs redundancy in the extracted syndrome.
Shor's method creates redundancy by repetition.
The straightforward implementation is to repeat extraction until the same syndromes are observed several times in a row
so that the probability of the observed syndromes being incorrect is sufficiently low \cite{Shor:1996}.

The point we make is that if well-chosen stabilizer operators are used in repetition,
the extracted syndrome in each repetition cycle need not be the same.
With the ability to detect incorrect syndrome bits,
we only need to consecutively observe coherent results that point to the same error on qubits until enough confidence is gained.

Moreover, if stabilizer operators are chosen so that most low-weight fault paths give distinct syndromes,
\textit{maximum likelihood decoding} \cite{MacKay:2003} or its approximation can be reliable enough
to infer a most likely fault path from a single extracted syndrome.
For instance, as was assumed in the previous sections, if syndrome extraction does not frequently introduce errors on data qubits,
it is reasonable to assume that the extracted syndrome bits are mostly correct
because the hypothetical ``correct'' syndrome does not change during the extraction process.
As we have seen, a good choice of generators or a few redundant stabilizer operators can be enough
to make the syndromes of likely errors all distinct under this error model.

It should be noted, however, that depending on the error model, tolerance against a decent number of erroneous syndrome bits may not be sufficient
to achieve the highest possible reliability.
For instance, if syndrome extraction itself likely causes quantum errors that drastically change what the correct syndrome should be,
a low-weight fault path can correspond to a large number of flips in the extracted syndrome.

To see how a newly introduced error on a single data qubit can cause a catastrophic effect,
take the four independent generators
\begin{align*}
S_0 &= XZZXI, & S_1 &= IXZZX,\\
S_2 &= XIXZZ, & S_3 &= ZXIXZ,
\end{align*}
of the perfect $5$-qubit code we used in Section \ref{sec:review}.
Assume that syndrome bits are extracted in order from $s_0$ to $s_3$ according to their subscripts.
It is a benign fault if the measurement of $S_0$ introduces, say, $Z$ on the fifth data qubit because the subsequent measurements will pick up on it and correct the error
as long as there was no error at the start of syndrome extraction and the procedure finishes otherwise perfectly.
However, if the final measurement involving $S_3$ introduces $Z$ on the first data qubit,
even if everything else is completed perfectly as intended, the two syndrome bits $s_0$ and $s_2$ are now ``wrong''
because the commutativity of $S_0$ and $S_2$ with the current error on data qubits is flipped due to $X$ on the first data qubit.
Even if we use the redundant stabilizer operator $S_4=\prod_{i=0}^{3}S_i$ as in Section \ref{subsec:single},
this error will slip through this round of syndrome extraction and should be identified during subsequent rounds.

Another example is failure of a controlled NOT (CNOT) gate between a data qubit and ancilla qubit that results in a double error, such as the back action of the CNOT gate.
This type of error can flip a syndrome bit while introducing a single error on data qubits.
Even if the rest of the quantum circuit operates perfectly,
the extracted syndrome of weight $1$ generally points to an error that is different from what is happening on data qubits.

It is notable that, with the help of $S_4$, the global $1$-error-correcting property may be able to detect the double error we just described.
For instance, if the error model is such that this type of error is fairly frequent compared to other kinds,
a reasonable inference algorithm would report this fault path of weight $1$ as a likely suspect, perhaps along with a single syndrome bit flip as another likely possibility.
If the next round of syndrome extraction finishes without an error, it will point to the former possibility rather than a hiccup on one syndrome bit during the first round,
giving the decoder a stronger clue about the error than if $S_4$ is not used.

As the above discussion shows, while it is generally beneficial to be able to correct erroneous syndrome bits or give more clues about the nature of noise,
it requires a sophisticated analysis to truly optimize the choice of stabilizer operators to a realistic error model for fault tolerance.

\section{Concluding Remarks}\label{sc:cr}
We have examined stabilizer quantum error correction and revealed its built-in tolerance against imperfect syndromes.
A challenging problem arose regarding optimizing the choice of stabilizer operators for a realistic error model.
Nevertheless, we were able to generalize Shor's syndrome extraction and opened a path to unlocking the hidden potential of stabilizer codes.
Indeed, we demonstrated that extra reliability may come at little or no cost by carefully choosing generators
in the sense that a stabilizer code can acquire error correction power for imperfect syndromes without increasing the number of physical qubits,
reducing the amount of encoded quantum information, or requiring many additional measurements.

An interesting question is when and how an $[[n,k,d]]$ stabilizer code can identify all likely fault paths through just $n-k$ independent generators.
From our observations, it appears that for given $n$ and $k$,
a stabilizer code with poorer distance parameter $d$ tends to possess a greater potential in correcting syndrome bits
because such a code leaves plenty of room in the available syndrome patterns for syndrome error correction.

Another important direction of research is how to optimize the choice of stabilizer operators in the context of fault-tolerant quantum computation.
In fault-tolerant syndrome extraction, the performance is affected also by many factors other than the maximum weight of errors a code can tolerate.
Ultimately, we would like to choose stabilizer operators in such a way that the chosen set unlikely introduces difficult errors, is the easiest to implement,
and leads to the best possible raw error correction power from the coding theoretic viewpoint.
While this is a very challenging problem, it is a very important one to be settled.

In particular, one of the remaining problems that deserve greater attention is
that the chosen generators and/or few extra stabilizer operators that are coding theoretically promising may not always be of low weight.
In many cases, it is important to use low-weight stabilizer operators for practical reasons.
Moreover, if the low weight propaty can not be guaranteed, it is plausible that Knill's and Steane's syndrome extraction can work better
than the idea of redundant syndromes in practice
as long as the implemented quantum error-correcting code is compatible with them.
Therefore, it is of importance to consider additional constraints that arise in practical situations.

We have made progress in robust syndrome extraction through a coding theoretic approach.
Nonetheless, this is just an initial step towards more general and realistic solutions.
As the feasibility of universal quantum computation rests on the shoulders of inevitably imperfect quantum error correction,
it is hoped that further progress will be made in this field.

%

\end{document}